\documentclass[]{jfm}
\usepackage[latin1]{inputenc}
\usepackage{graphicx}
\usepackage{amsmath}
\usepackage{amssymb}
\usepackage{color}
\usepackage[dvips]{epsfig}
\usepackage{epstopdf}
\usepackage{times,mathptmx}
\usepackage{epstopdf}
\usepackage{natbib}

\newcommand{\eps}{\varepsilon}
\newcommand{\Ca}{\mathrm{Ca}}

\begin{document}
\shortauthor{E. S. Asmolov et al} 
\title{Flow-driven collapse of lubricant-infused surfaces}

\author[E. S. Asmolov et al]
{Evgeny S. Asmolov$^{1}$, Tatiana V. Nizkaya$^{1}$ and Olga I. Vinogradova$^{1,2}$
\thanks{Author to whom correspondence should be addressed; email: oivinograd@yahoo.com}}

\affiliation{
$^1$A.N.~Frumkin Institute of Physical Chemistry and Electrochemistry,\\[\affilskip]
Russian Academy of Sciences, 31 Leninsky Prospect, 119071 Moscow, Russia\\[\affilskip]
$^2$DWI - Leibniz Institute for Interactive Materials, Forckenbeckstr. 50, 52056 Aachen, Germany
}
\date{Received: date / Accepted: date}

\maketitle

\begin{abstract}
Lubricant-infused surfaces in an outer liquid flow generally reduce viscous drag. However, owing to the meniscus deformation,  the infused state could collapse. Here we discuss the transition between infused and collapsed states of transverse shallow grooves, considering the capillary number, liquid/lubricant viscosity ratio and aspect ratio of the groove as parameters for inducing this transition. It is found that depending on the depth of the grooves, two different scenarios occur. A collapse of lubricant-infused surfaces could happen due to a depinning of the meniscus from the front groove edge. However, for very shallow textures, the meniscus contacts the bottom wall before such a  depinning could occur. Our interpretation could help  avoiding this generally  detrimental effect in various applications.
\end{abstract}
\section{Introduction}

Slippery lubricant-infused surfaces  have received much attention in recent years since they provide a drag reduction and flow manipulation in microfluidic devices~\citep{wong.ts:2011,nizkaya.tv:2014,solomon.br:2014,keiser.a:2017}.  The lubricant could be a gas trapped by superhydrophobic (SH) textures or another liquid, such as oil. SH surfaces show very large effective slip length~\citep{ybert.c:2007,vinogradova.oi:2011}, which makes them attractive for use in microfluidic applications~\citep{vinogradova.oi:2012}. Liquid-infused (LI) surfaces are less slippery~\citep{asmolov.es:2018}, but are commonly considered to be potentially more stable and robust against pressure-induced failure compared to SH surfaces, which makes them useful in various applications, including anti-biofouling~\citep{epstein.ak:2012} and ice-phobicity~\citep{kim.p:2012}.
Implementation of LI surfaces often requires a thorough understanding of the dynamics of a
lubricant within a patterned substrate that is
exposed to external hydrodynamic flow. This fundamental problem also applies to a variety of similar
situations, including the stability of small bubbles or droplets, trapped by slightly rough or heterogeneous surfaces~\citep{vinogradova.oi:1995b,borkent.b:2007}.

The influence of a curved meniscus on the slipping properties of SH texture has been extensively studied analytically \citep{sbragaglia.m:2007}, numerically \citep{teo.sj:2010} and in experiments \citep{karatay.e:2013,Xue2015}.  In all these studies protrusion or inflection of the meniscus has been achieved by changing the hydrostatic pressure in gas, but the flow-induced dynamic deformations of the meniscus has been neglected. Deformation of liquid/gas interface by the flow has been studied only for strongly protruding bubbles \citep{gao2009,harting.j:2008}. In particular, simulation studies \citep{harting.j:2008} have shown that at large capillary numbers pinned surface bubbles are deformed by an external viscous flow, which dramatically alters the slip length of the SH texture, but no attempt has been made to address the issue of their stability. We are also unaware of any study of the dynamic deformation of an initially flat meniscus.

Existing theories describing a stability of lubricant-infused state mostly include the
configurations of static wetting drops at the SH surface. Several static criteria have
been suggested~\citep{cottin-bizonne2004,bico.j:2002}, and later extended to a more complex, metastable situations~\citep{reyssat.m:2008,dubov.al:2015}. The body of theoretical and experimental work investigating the stability of SH and LI surfaces in external flows is rather scarce, although there exists some recent literature in the area. ~\citet{wexler.js:2015,Liu2016} have considered an outer shear flow aligned with the direction of extended closed grooves, where a reverse pressure gradient in a lubricant is generated. As a result, a curvature of the static meniscus is largest near the channel inlet, and the failure of deep LI grooves occurs when the dynamic contact angle becomes large, while that of shallow grooves - when the meniscus contacts the groove bottom. The collapse of partially filled deep SH and LI grooves induced by an external transverse shear has been studied numerically by~\citet{ge2018}. These authors concluded that the  induced by such a flow meniscus deformation decreases with the lubricant/outer liquid viscosity ratio and that the collapse of lubricant-infused grooves is possible only when this ratio is smaller than unity.


In this paper we present some results of a study of the possible collapse of shallow lubricant-infused grooves driven by an external transverse shear flow. Our model, which is different from configurations explored before, assumes that the meniscus is initially flat and pinned at the groove edges, and that grooves are unbounded in the longitudinal direction. To solve a two-phase problem we couple a general solution of Stokes problem for an outer flow and a lubrication approach for flow in a groove. Note that similar strategy has previously been successfully employed for investigating flows over flat SH textures \citep{maynes1,nizkaya.tv:2013} and of evaporating thin film \citep{doumenc2013}. We shall see that flow, induced in a lubricant layer, strongly depends on its local thickness, which is in turn controlled by a local pressure gradient, and that the stationary shape of a deformed meniscus becomes roughly antisymmetric (concave-convex). The meniscus depinning from the front groove edge occurs when the advancing contact angle for an outer liquid is reached at some critical capillary number. However, very shallow lubricant-infused textures collapse when the deformed meniscus contacts the groove bottom. One of important differences between our results and prior work, that employed different
models, is that the lubricant-infused surface becomes more stable when the lubricant/liquid viscosity ratio is smaller. Therefore, contrary to a common belief for some geometries SH surfaces could be  more robust than LI ones.

The paper is arranged as follows. In Sec.~\ref{geqs} we describe the model of a lubricant-infused shallow groove and derive asymptotic equations for a two-phase flow problem. Sec.~\ref{res} contains results of our numerical calculations. We conclude in Sec.~\ref{concl}. The calculation details of the meniscus shape and of the effective slip length are given in Appendix~\ref{calc}, and we justify the use of the lubrication approximation for an inner flow in Appendix \ref{valid}.

\section{Asymptotic theory}
\label{geqs}

\subsection{Model}
\begin{figure}
\centering
\includegraphics[width=9cm]{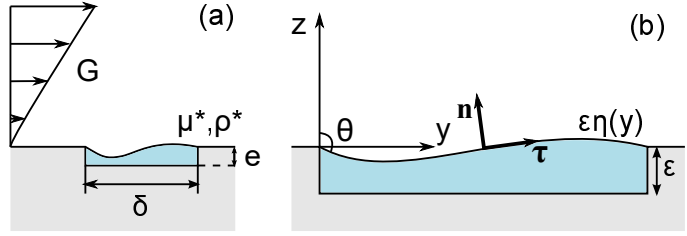}
    \caption{Sketches of (a) an outer shear flow past a shallow lubricant-infused groove of width $\delta$ and depth $e$, and of (b) liquid/lubricant interface $\eps \eta(y)$ in dimensionless coordinates. The concave liquid/lubricant interface meets the front groove edge with an angle $\theta$ defined relative to a vertical.}
  \label{sketch}
\end{figure}

\begin{figure}
\centering
\includegraphics[width=8cm]{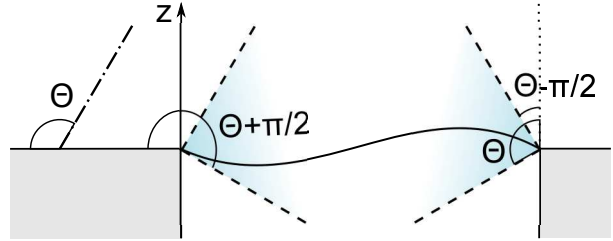}
    \caption{Pinning of a contact line on a rectangular groove edge. The liquid meets the solid
with a contact angle $\Theta$. Hence the contact angle at the groove edges can take any value (if the horizontal direction is
considered as the reference one) between $\Theta$ and $ \Theta + \pi/2$, as illustrated by colored region confined between dashed lines. The attainable contact angles at the edge, redefined relative to $z$-direction, are then confined between $\Theta - \pi/2$  and $\Theta$.}
  \label{sketch_depin}
\end{figure}

We consider a linear shear flow of an outer liquid of viscosity $\mu^*$ over a shallow rectangular groove of width $\delta$ and depth $e\ll \delta$ (see Fig.\ref{sketch}(a)), filled with a lubricant of viscosity $\mu^{l*}$ (hereafter the asterisk denotes dimensional variables).
  The $y^*$-axis is aligned with the shear direction, while the $z^*$-axis is defined normal to the wall, and the dimensionless coordinates are introduced as $y=y^*/\delta$ and  $z=z^*/\delta$. We assume that the groove aspect ratio is small, $\eps=e/\delta\ll 1$, and that at rest the static liquid/lubricant interface is flat and located at $z=0$, i.e. there is no pressure difference across the meniscus that is pinned at the edges of the groove. When the meniscus is perturbed by an external flow, its slightly deformed shape sketched in Fig.\ref{sketch}(b) can be described
  by a function $\eps \eta(y)$, where $\eta=O(1)$.
  It is convenient to introduce  the stationary angle $\theta \geq \pi/2$  defined (relative to a vertical direction) at the point where the concave or flat meniscus meets the groove edge (here the front one). Thanks to the lubricant volume conservation, the meniscus deformation should be rather close to antisymmetric, so that the stationary angle at the opposite grove edge (here the rear one) is about $\pi - \theta$. The observed angle $\theta$ cannot exceed a limiting value known as an advancing angle, beyond which the contact line  does depin from the groove edge and  move. Likewise, when $\pi - \theta$ decreases down to a limiting value of the receding angle, the contact line should suddenly shift laterally.

  If we consider a chemically homogeneous (ideal) surface, the bounds of attainable values of $\theta$ are determined univocally by a liquid contact angle $\Theta$ (above $\pi/2$ to  provide a lubricant-infused state) on a planar horizontal surface (see Fig.\ref{sketch_depin}). Its value can, of course, always be adjusted by a suitable modification of the solid surface \citep{grate2012correlation, dubov.al:2015,jung2009wetting}, but note that on the most solids $\Theta$ never exceeds $2 \pi /3$ or $120^{\circ}$ \citep{wexler.js:2015,yakubov2000contact,jung2009wetting}.
    Following \citet{quere.d:2008,herminghaus2008wetting,dubov2018boundary} one can argue that  the contact angle of a liquid at the groove edges can be any between $\Theta$ (receding) and $\Theta + \pi/2$ (advancing) as illustrated in Fig.\ref{sketch_depin} by the colored regions that are symmetric relative to a midplane of the groove. The attainable contact angles redefined relative to $z$-direction are then confined between $\Theta - \pi/2$ (receding)  and $\Theta$ (advancing). We remark and stress, however, that the bounds of attainable angles are asymmetric relative to $z=0$ except the case of $\Theta = 3 \pi /4$ (or $135^{\circ}$). Of the two possible unstable angles, the one is normally attained faster and overshadows the other. Since for our surfaces $\Theta$ is smaller than $3 \pi /4$, one can argue that
  the depinning will occur on that edge of the groove, where the meniscus is concave, and when $\theta$ will reach the value of $\Theta$.
    For a meniscus shape sketched in Figs. \ref{sketch} and \ref{sketch_depin} this would be the front (left) edge, but of course such a shape is by no means obvious,  and will be  justified below  by solving a hydrodynamic problem.
  As a side note we mention that the depinning would occur simultaneously at front and rear (right) edges if $\Theta = 3 \pi /4$, and for larger $\Theta$ - on a rear edge, when $\theta \simeq \Theta - \pi/2$. Clearly, these estimates hold only for ideal surfaces and relatively low speed. They would become approximate when speed is large enough, and dynamic angle would deviate from $\Theta$, or when surface is chemically heterogeneous. However, they provide us with some guidance.

\begin{figure}
\centering
\includegraphics[width=10cm]{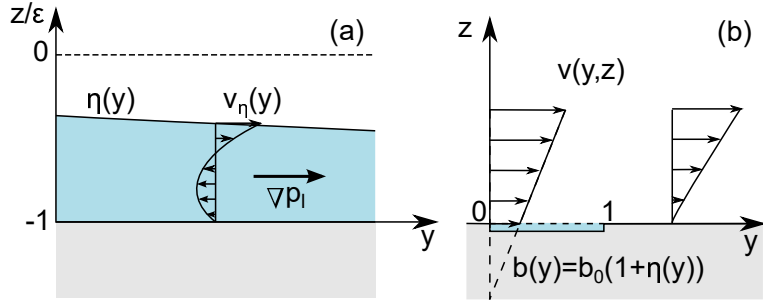}
    \caption{(a) Recirculation flow in a shallow groove. (b) Shear flow and a local slip length in outer fluid.}
  \label{in_flow}
\end{figure}

The dimensionless velocity and pressure are defined as $\mathbf{u}=\mathbf{u}^*/(G\delta)$ and $p=p^*/(G\mu^*)$, where $G$ is an undisturbed shear rate. We stress, that near the groove the outer flow is modified due to a slippage at the liquid/lubricant interface, and that the lubricant flow, induced by a reverse pressure gradient, has zero flow rate in any cross-section (see Fig. \ref{in_flow} (a)). The flows in a lubricant and an outer liquid are stationary and satisfy Stokes equations


\begin{eqnarray}
\mathbf{\nabla }\cdot \mathbf{u}=0,\quad \Delta \mathbf{u-\nabla }p=\mathbf{0%
},  \label{Stokes}
\\
\mathbf{\nabla }\cdot \mathbf{u}^{l}=0,\quad \mu \Delta \mathbf{u}^{l}%
\mathbf{-\nabla }p^{l}=\mathbf{0},
\label{Stokes_l}
\end{eqnarray}%
where $\mathbf{u}=(0,v,w)$ and $\mathbf{u}^{l}=(0,v^l,w^l)$ are velocity fields in liquid and in lubricant, $p, p^l$ are corresponding pressure distributions, and $\mu=\mu^{l*}/\mu^*$ is the lubricant/liquid viscosity ratio.
Far from the lubricant-induced surface the liquid flow represents a linear shear flow,
$\mathbf{u}|_{z\to\infty}=z\mathbf{e}_y$,
where $\mathbf{e}_y$ is a unit vector along $y-$axis. We apply no-slip condition at solid boundaries, and at the liquid/lubricant interface, $\eps\eta(y)$, we use the conditions of impermeability
\begin{equation}
 \mathbf{u\cdot n=u}^{l}\mathbf{\cdot n}=0,  \label{imper}
\end{equation}%
and of continuity of tangent velocity and tangent stresses,
\begin{eqnarray}
 \mathbf{u\cdot \tau }=\mathbf{u}^{l}\mathbf{\cdot \tau },  \label{covel}\\
\mathbf{\tau \cdot \sigma }\cdot \mathbf{n}=\mathbf{\tau \cdot \sigma }%
^{l}\cdot \mathbf{n},
\label{shear_str}
\end{eqnarray}%
where $\mathbf{\tau}$ and $\mathbf{n}$ are unit tangent and outward normal (to the meniscus) vectors,  and the stress tensors are $\mathbf{\sigma =%
\mathbf{\nabla u+}}\left( \mathbf{\mathbf{\nabla u}}\right) ^{T}\mathbf{-}p%
\mathbf{I}$ and $\mathbf{\sigma}^l =\mu\left(
\mathbf{\nabla u}^l+( \mathbf{\nabla u}^l) ^{T}\right)\mathbf{-}p^l%
\mathbf{I}$.

The condition for normal stresses at the interface can be derived using the Laplace equation,
\begin{equation}
 \mathbf{n}\cdot (\sigma^{l}-\sigma)\cdot \mathbf{n}=\frac{\kappa}{\mathrm{Ca}},
\label{laplace}
\end{equation}%
where $\Ca=\mu ^{\ast }G\delta/\gamma $ is the capillary number defined using an outer liquid viscosity and $\kappa \simeq \eps \eta''$
is the interface curvature (negative for the meniscus protruding into the outer liquid).

These equations should be supplemented by the condition of volume conservation in the lubricant phase and by pinning conditions at the edges of the groove,
\begin{equation}
\int\limits_{0}^{1}\eta(y)dy=0, \quad \eta(0)=\eta(1)=0.
\label{pinning}
\end{equation}%

Equations (\ref{Stokes})-(\ref{pinning}) represent a closed system governing liquid and lubricant flows. They involve three dimensionless parameters, i.e. the shallow groove aspect ratio $\eps$, and lubricant/liquid viscosity ratio $\mu$, and the capillary number $\Ca$. The two latter parameters could be any, but $\eps$ is small, so that we could use it to construct asymptotic solutions for the meniscus shape, velocity fields, and pressure.

Since $\eps$ is small,  $\mathbf{\tau }\simeq \left(
0,1,\eps\partial _{y}\eta \right)$ and $\mathbf{n}\simeq \left( 0,-\eps\partial _{y}\eta ,1\right)$, and boundary conditions (\ref{imper})-(\ref{shear_str}) at a curved interface can be simplified to
\begin{equation}  w=w^l=0, \quad v=v^l=v_{\eta}(y),
\quad
\dfrac{\partial v}{\partial z}=\mu\dfrac{\partial v^l}{\partial z}, \label{interface_eps}
\end{equation}%
which define a coupling between a liquid and a lubricant. Here
$v_{\eta}(y)$ is the tangent velocity of both liquid and lubricant at the interface.
 From  (\ref{laplace}) we then obtain
\begin{equation}
 \sigma^l_{zz}=\sigma_{zz}+\dfrac{\eps\eta''}{\Ca},\label{norm_stress}
\end{equation}
which determines the meniscus shape, with normal stresses given by
\begin{equation}
\sigma_{zz}=2 \partial _{z}w-p,\quad \sigma^l_{zz}=2\mu \partial _{z}w^{l}-p^{l}.
\label{s_zz}\end{equation}
Note that the normal stresses include not only the pressure but also gradients of normal velocities since unlike the no-slip case, the latter do not vanish at slippery surfaces.



\subsection{Inner flow  }

The lubricant flow in the groove is generated by the interface velocity $v_{\eta}(y)$.  For this inner problem the boundary condition,  Eq.(\ref{interface_eps}), should be  imposed at the curved meniscus $z=\eta(y)$, since its local deviation from the flat one, $z=0$, is comparable to the depth of the groove, $\eps$. Since the local slopes of the meniscus are small, we apply the lubrication theory and consider a locally parabolic velocity profile of zero flow rate:
\begin{equation}
v^l\simeq v_{\eta }\left( y\right) \zeta\left(3\zeta-2\right)
,\quad w^l\simeq0,
\label{v_l}
\end{equation}%
 where $\zeta=(z+\eps)/[\eps(1+\eta)]$ varies from $0$ at the bottom wall to $1$ at the interface. Eqs.\eqref{v_l}, which are equivalent to derived by \citet{nizkaya.tv:2013} for a flat meniscus, but varying local thickness of the thin lubricant film, allow one to immediately calculate both the lubricant shear rate at the interface
\begin{equation}\partial_z v^l=\dfrac{4 v_{\eta}}{\eps(1+\eta)}.\end{equation}
and the transverse local slip length
\begin{equation}
b(y)=\frac{v_{\eta}}{\mu \partial_z v^l} \simeq \frac{\eps}{4 \mu}(1+\eta),
\label{by}
\end{equation}
where $\eps/\mu=O(1)$. Eq.\eqref{by} implies that $b(y)$ is proportional to a local thickness of the lubricant layer and inversely proportional to $\mu$, similarly to infinite systems~\citep{vinogradova.oi:1995a,miksis.mj:1994}. We should also note that $\dfrac{\eps}{4 \mu}$ may be interpreted as a local transverse slip length, $b_0$, on a flat (undisturbed) liquid/lubricant interface~\citep{nizkaya.tv:2013,nizkaya.tv:2014}.

 For SH grooves, pressure induced by flow changes in the inner gas is usually neglected since $\mu \ll 1$. However, $\mu$ takes on finite values for lubricant-infused surfaces, and the gradient of pressure in closed shallow grooves could become very large, even when the lubricant viscosity is small, $\mu\sim\eps$. Using simple scaling arguments one can show that $\partial _{y}p^{l} \simeq \mu\dfrac{\partial^2v^l}{\partial y^2}\sim \mu \eps^{-2}\gg 1$. Indeed,  the corresponding to Eq.(\ref{v_l}) pressure profile satisfies
\begin{equation}
 \partial _{y}p^{l}\simeq \dfrac{6\mu v_{\eta }\left( y\right) }{%
\eps^{2}\left( 1+\eta \right) ^{2}},\quad \partial _{z}p^{l}\simeq 0.\label{p_l}
 \end{equation}

Finally, using \eqref{p_l} and \eqref{Stokes_l} one  can estimate that
$2\mu\partial_z w^l \simeq -2\mu\partial_y v^l\sim \mu\ll |p^l|$. From Eqs.\eqref{s_zz} it then follows that $\sigma^l_{zz} \simeq -p^{l}$, which may be determined by integrating Eq.(\ref{p_l}).

\subsection{Outer flow}

We now turn to an outer liquid flow (of length scale $\delta$), that is practically cannot be affected by small variations in $\eps\eta$, and, therefore, Eqs. (\ref{interface_eps}) could be mapped
to the flat interface, $\left. \mathbf{u}\right\vert _{z=\eps\eta }=
\left. \mathbf{u}\right\vert _{z=0}+O(\eps).$
Impermeability condition Eq.(\ref{imper}) is then reduced to $w(y,0)=0,$ and Eq.(\ref{shear_str}) takes the form of a conventional partial slip condition applied at $z=0$,
 \begin{equation}
v = b(y)\dfrac{\partial v}{\partial z},
\label{slip0}
\end{equation}
with the local slip length described by \eqref{by}.

An outer solution for a flow field with prescribed velocity $v_{\eta
}\left( y\right) $ and zero normal velocity can be found using $u(y,z)$ for a
longitudinal configuration \citep{asmolov:2012}:%
\begin{gather}
v=u+z\frac{\partial u}{\partial z},\quad w=-z\frac{\partial u}{\partial y},
\label{vw} \\
p=-2\frac{\partial u}{\partial y}.  \label{p}
\end{gather}%
Here $u$ satisfies the Laplace equation, $\Delta u=0$, with the boundary condition $%
u\left( 0,y\right) =v_{\eta }\left( y\right)$. Eqs.~(\ref{vw}), (\ref{p}) immediately suggest that $\sigma
_{zz}\left( 0,y\right)=2 \partial _{z}w-p=0$. In other words, the contributions of the pressure and the gradient of normal velocity to the normal stress cancel out. This implies that the outer flow does not affect the meniscus deformation.
Consequently, the meniscus shape depends on the sign of the pressure gradient in lubricant. When it is positive, the shape will be as shown in Figs. \ref{sketch} and \ref{sketch_depin}, i.e. concave at the front groove edge and convex at the rear one.

The equation describing the meniscus shape can be obtained by differentiating Eq.(\ref{norm_stress}) with respect to $y$. Keeping then only the leading term in $\eps$ and using Eq.(\ref{v_l}) we find that this shape obeys
\begin{equation}
\eta'''=-\frac{6 \mu \Ca}{\eps^3}\dfrac{v_{\eta}(y)}{(1+\eta)^2}.
\label{meniscus}
\end{equation}
To solve this differential equation, conditions \eqref{pinning} should be imposed.

To summarize, the outer asymptotic problem is reduced to Eq.(\ref{Stokes}) coupled with the equation for  the meniscus shape (\ref{meniscus}) expressed via the local slip length $b(y)$, Eq.(\ref{by}), and interface velocity $v_{\eta }$.

\subsection{Limiting cases}

In the general case, the inner and outer flows are strongly coupled, and the two-phase problem should be solved numerically. However, in some limits the system could be simplified, thanks to a decoupling of  these flows.

In the limit of $\eps/\mu\ll 1$, typical for very viscous lubricant and/or extremely thin lubricant layer, Eq.(\ref{slip0}) reduces to a no-slip boundary condition, $v(y,0)\simeq 0$. Consequently, an  outer flow remains undisturbed by the inner one, and the shear stress in liquid is $\partial _{z}v(y,0) \simeq 1$. It follows then from \eqref{interface_eps} that the  lubricant shear rate is  $\partial _{z}v^{l}(y,0) \simeq \mu ^{-1}$, and the interface velocity, found from Eq.(\ref{slip0}), is  $v_{\eta }=b(y)\partial _{z}v(y,0)\simeq \eps(1+\eta)/4 \mu $, i.e. it decreases with $\mu$. From Eq.\eqref{p_l} it follows then that $\partial _{y}p^{l}$ is finite and does not depend on $\mu$, and Eq.(\ref{meniscus}) reduces to
\begin{equation}
\eta''' =-\dfrac{3 \Ca }{2 \eps^2 \left( 1+\eta \right) }.
\label{meniscus0}
\end{equation}

In the opposite limiting case of low lubricant viscosity, $\eps/\mu \gg 1$, the local slip length $b(y)$ is large
provided $1+\eta (y)$ remains finite. Thus we might argue that at any deformation a sensible approximation for a local slip would be $b(y) \rightarrow \infty$ that leads to the interface
velocity $v_{\eta }^{P}\left( y\right)=\frac{1}{4}\left[1-\left(2y-1\right)^2\right]^{1/2}$\citep{philip.jr:1972}. Substituting this to Eq.(\ref{meniscus})  we get
\begin{equation}
\eta'''=-\frac{6 \mu \Ca}{\eps^3}\dfrac{v_{\eta}^p(y)}{(1+\eta)^2} .
\label{meniscus_inf}
\end{equation}

\section{Results of calculations and discussion}
\label{res}

In this section we present some numerical results for a model system formulated above. The details of our calculations are presented in Appendix~\ref{calc}. In Appendix~\ref{valid} we present some numerical results justifying the use of the lubrication approximation.

\begin{figure}
\centering
\includegraphics[height=4cm]{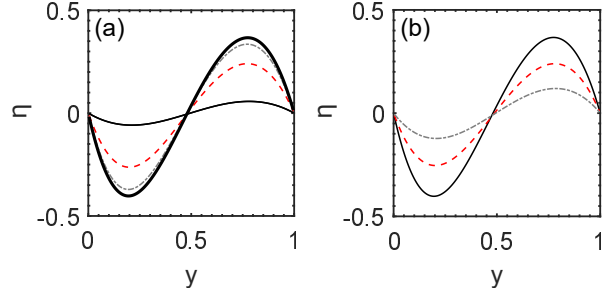}
    \caption{(a) Meniscus shape computed at $\eps=0.1$ and $\Ca=0.3$. Solid, dashed and dash-dotted curves show results obtained using $\mu = 0.02$, $0.2$ and $1$.  Bold curve shows calculations from asymptotic Eq.(\ref{meniscus0}). (b) Predictions of Eq.(\ref{meniscus0}) for $\eps=0.1$ and $\Ca=0.1$, $0.2$, $0.3$ (dash-dotted, dashed and solid curves).}
      \label{eta_and_p}
\end{figure}

We have first investigated the dynamic meniscus stationary shape, $\eta(y)$, at fixed $\eps=0.1$. Fig.~\ref{eta_and_p}(a) shows the numerical results obtained using $\Ca=0.3$, and several typical viscosity contrasts, $\mu=0.02$ (water/air interface), $\mu=0.2$ (oil/water), and $\mu=1$, where liquid and lubricant are of the same viscosities. Our calculations confirm that the function $\eta(y)$ is nearly antisymmetric and has two extrema. It takes its minimum value close to the front edge of the groove (region of concave curvature), and a maximum occurs in the vicinity of the rear edge, where $\eta(y)$ inverts its curvature to convex. As predicted and discussed above, this implies that that the depinning occurs at the front edge.   The absolute values of extrema decrease with $\eps/\mu$, which implies that with our parameters the meniscus deformation grows with $\mu$. Also included are predictions of asymptotic Eq.(\ref{meniscus0}), which determines an upper bound on meniscus deformation, attainable for very small $\eps/\mu$. For smaller values of $\Ca$, the maximum possible meniscus deformation decreases as illustrated in Fig.~\ref{eta_and_p}(b). We also note that a larger deflection of the meniscus shape from the flat one is always accompanied by an increase in $\theta = \pi/2-\arctan(\eps\eta'(0))$.

\begin{figure}
\centering
\includegraphics[height=4cm]{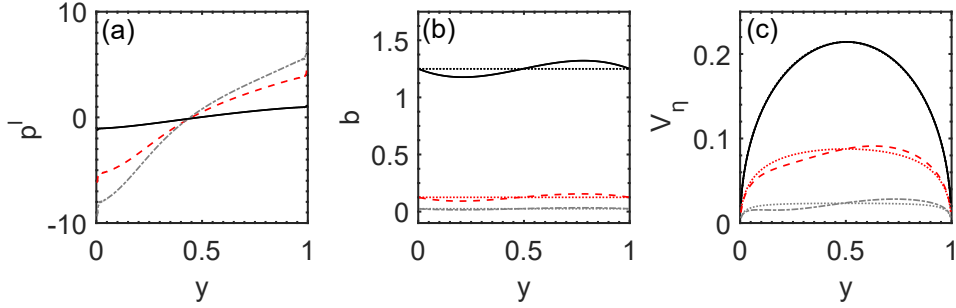}
    \caption{The lubricant pressure (a),  local slip length (b), and  interface velocity (c) calculated using $\eps=0.1$ and $\Ca=0.3$. Solid, dashed and dash-dotted curves show results for $\mu = 0.02$, $0.2$ and $1$. Dotted lines show results for a flat meniscus.}
  \label{b_and_V}
\end{figure}

As described in Sec.~\ref{geqs}, the local curvature of the meniscus is associated with pressure in the lubricant film, which in turn can be related to the lubricant flow. Fig. \ref{b_and_V} plots the lubricant pressure, local slip length  $b(y)$ and the interface velocity $v_{\eta}(y)$ computed with the same parameters as in Fig.~\ref{eta_and_p}(a). We see that $p^l$ increases with $y$, and its (positive) gradient is smaller for larger $\eps/\mu$, that implies that at fixed $\eps$ pressure gradient grows with $\mu$. By contrast, both $b(y)$ and $v_{\eta}(y)$ increase with $\eps/\mu$. The numerical data also show that at large $\eps/\mu$ the interface velocity remains close to  $v_{\eta}^P(y)$, but for smaller slip lengths it is significantly affected by the deformation of a lubricant film.

\begin{figure}
\centering
\includegraphics[height=4cm]{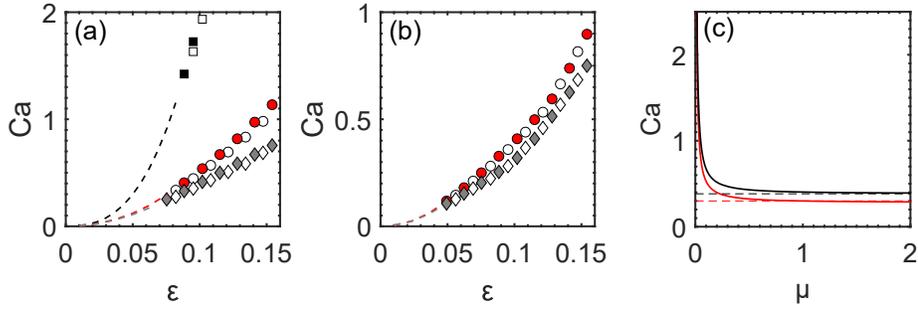}
\caption{(a) Critical values of $\Ca$, beyond which the lubricant-infused surfaces collapse, versus $\eps$ calculated using $\Theta=120^\circ$ and $\mu=0.02$ (squares), $0.2$ (circles), $1$ (diamonds). Filled and open symbols are obtained from Eqs.\eqref{meniscus} and \eqref{meniscus1}, and correspond to the depinning. Dashed curves correspond to a contact of the liquid/lubricant interface with the bottom wall.  (b) The same, but for $\Theta=110^\circ$.   (c) Critical $\Ca$ at which the depinning occurs as a function of $\mu$ calculated for $\eps=0.1$. Solid curves from top to bottom show results obtained using $\Theta=120^\circ$ and $110^\circ$. Dashed lines plot the corresponding asymptotic values calculated from Eq.(\ref{meniscus0}). }
 \label{diagram}
\end{figure}

To examine the scenario of a collapse more closely, the critical $\Ca$ has been calculated as a function of $\eps$  for several  values of $\mu$ (taken the same as in Figs.~\ref{eta_and_p} and \ref{b_and_V}). Specimen results obtained
using $\Theta = 120^\circ$ and $110^\circ$ that are close to those observed experimentally \citep{grate2012correlation, dubov.al:2015,wexler.js:2015,jung2009wetting} are included in Figs.~\ref{diagram}(a) and (b), where we denote by filled symbols in the $(\Ca, \eps)$ plane the values of $\Ca$, which correspond to $\theta = \Theta$. It is well seen that for smaller $\Theta$ the depinning should occur at smaller $\Ca$. Note that these results well agree with calculations made using the exact equation for the meniscus curvature (shown by open symbols), confirming the validity of our approximations.
     As $\mu$ increases the required for a depinning value of $\Ca$ reduces, and for sufficiently large $\mu$ should approach the value calculated from Eq.(\ref{meniscus0}). The curve for the critical $\Ca$ of depinning as a function of $\mu$, calculated using $\eps=0.1$, is included in Fig.~\ref{diagram}(c). It can be seen that it reduces rapidly at small viscosity contrast and saturates to a constant value given by Eq.(\ref{meniscus0}) already at $\mu \geq 1$. We emphasize that at smaller $\eps$ our non-linear system has no positive stationary solution when $\Ca$ becomes larger than some critical value. As $\Ca$ approaches this value, the local thickness of a lubricant film tends to zero, which is accompanied by an infinite growth of the pressure gradient in the film neck (see Appendix~\ref{calc}), but note that $\theta$ still remains smaller than $\Theta$. Therefore,
     one might argue that for sufficiently small $\eps$ the curve of failure of lubricant-infused surfaces included in Figs.~\ref{diagram}(a) and (b) reflects a contact of a deformed meniscus with the bottom wall that occurs before the value of $\Theta$ is reached. It is interesting that the curves corresponding to these two scenarios of collapse of lubricant-infused surfaces meet smoothly at $\eps \simeq 0.07$ ($\Theta = 120^\circ$) and $\eps \simeq 0.05$ ($\Theta = 110^\circ$) in Figs.~\ref{diagram}(a) and (b).


\begin{figure}
\centering
\includegraphics[height=4cm]{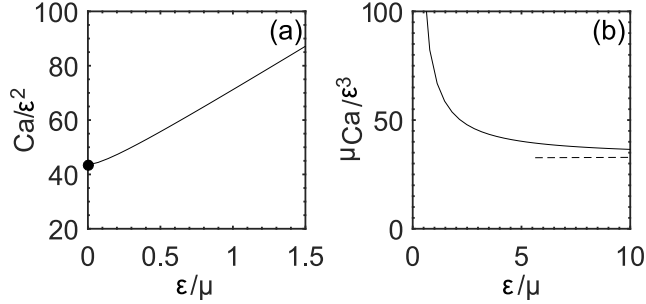}
    \caption{(a) Critical $\Ca /\eps^2$ (solid curve), corresponding to a contact of the meniscus with the bottom wall, as a function of $\eps / \mu$. The solution of Eq.(\ref{meniscus0}) is shown by a filled circle. (b) Critical $\mu \Ca /\eps^3$ (solid curve) and predictions of Eq.(\ref{meniscus_inf}) that determines the asymptotic slope of critical $\Ca /\eps^2$ at large $\eps/\mu$ (dashed line).}
  \label{critical_diagram}
\end{figure}

Finally, we recall that $b_0$ is proportional to $\eps / \mu$, so that based on Eq.(\ref{meniscus}) one can suggest that critical $\Ca /\eps^2$ is a universal scaled capillary number that allows one to evaluate when the meniscus contacts a bottom and this scenario of a failure of lubricant-infused surfaces occurs. To illustrate this we plot in Fig.~\ref{critical_diagram}(a) the curve separating liquid-infused and collapsed states of the lubricant-infused surfaces in the $(\Ca /\eps^2, \eps / \mu)$ plane. The stable configuration corresponds to an area under the curve, while in the upper region the lubricant-infused surface is never stable. At $\eps / \mu \to 0$ we recover a solution of Eq.(\ref{meniscus0}).
The critical $\Ca /\eps^2$ increases with $\eps / \mu$,  and at sufficiently large $\eps / \mu$ it grows practically linearly  with the slope $O(10)$. This is better seen in Fig.~\ref{critical_diagram}(b), where we reproduced our data in the $(\mu \Ca /\eps^3, \eps / \mu)$ plane. When $\eps / \mu \to \infty$, the solution of Eq.(\ref{meniscus_inf}) becomes exact, but we might argue that it becomes a sensible approximation when $\eps / \mu$ becomes large, i.e. for SH surfaces.

\section{Conclusion}
\label{concl}

We have studied the meniscus deformation in an outer shear flow oriented transverse to lubricant-infused shallow grooves. It has been shown that the deviations of meniscus shape from the initial one, are mostly controlled by the inner, pressure-driven, lubricant flow.
While such a deformation practically does not affect the value of the slip length, it could induce the collapse of the lubricant-infused surface. Whether or not such a collapse should occur depends on the capillary number $\Ca$, lubricant/liquid viscosity ratio $\mu$, and aspect ratio of the groove $\eps$. Our work has shown that unlike the previously considered case of deep grooves, for shallow grooves the meniscus deformation increases with $\mu$. The mechanism of the failure of lubricant-infused shallow grooves depends, in turn, on the value of $\eps$. We have identified two separate mechanisms of failure of lubricant-infused state of surfaces. This could happen due to a depinning of the meniscus from the front groove edge,  when the value of advancing contact angle is reached. However, in the case of very small $\eps$, the meniscus contacts the bottom wall before such a depinning could occur.

 We have already mentioned the prior numerical work of~\citet{ge2018} who studied  deep transverse grooves, filled by a lubricant only partly (implying the mobility of the contact line), and found  that the meniscus deformation decreases with $\mu$. One important difference of our results for shallow transverse grooves with pinned contact lines is that the deformation decreases upon reducing $\mu$, likewise in the case of deep longitudinal grooves~\citep{wexler.js:2015,Liu2016}. We are unaware of any reported measurements of failure of transverse LI grooved surfaces. It would be very timely to test  theoretical predictions for transverse LI grooves by experiments.
 
 Finally, we recall that here an ideal, chemically homogeneous, surface has been assumed. For such a surface, the advancing liquid contact angle (relative the vertical) is exactly equal to a (Young) angle $\Theta$. For a chemically heterogeneous material the advancing angle will, of course, exceed the value of $\Theta$, making the LI surface more stable against depinning. Conversely, a hypothetic surface of very large $\Theta$, where depinning would be expected at the rear groove edge, becomes less stable when chemically heterogeneous since the receding angle is smaller than $\Theta$.

\begin{acknowledgments}
This research was partly supported by the Russian Foundation for Basic Research (grant 18-01-00729) and the Ministry of Science and Higher Education of the Russian Federation.
\end{acknowledgments}

\appendix
\section{Numerical method}
\label{calc}

The asymptotic equations \eqref{meniscus0} and \eqref{meniscus_inf} are solved using Runge-Kutta procedure and Newton method to satisfy boundary conditions (\ref{pinning}).
In general case $\eps/\mu=O(1)$, Eqs.~\eqref{Stokes},~\eqref{slip0} and~\eqref{meniscus} are coupled and we use an iteration scheme, starting from some initial guess.
Once the meniscus shape on $k$-th iteration is known, we solve the Stokes equations in liquid Eq.\eqref{Stokes} with the local slip length $b^k(y)=b_0[1+\eta^k(y)]$. We calculate the outer flow for periodical grooves \citep{nizkaya.tv:2013}  and expand the solution into Fourier series on a computational domain with a period $L=5\delta$ (where the solution is no longer dependent on $L$). The local slip boundary conditions are satisfied using collocation method. The computed interface velocity  $v_{\eta}^k(y)$ is then substituted into the right-hand side of Eq.(\ref{meniscus}) to obtain the next iteration:
\begin{equation}
\dfrac{\partial^3\eta^{k+1}}{\partial y^3}=\frac{6 \mu \Ca}{\eps^3}\dfrac{v_{\eta}^k(y)}{(1+\eta^k)^2}.
\label{men_iter}
\end{equation}
The meniscus shape is also sought in terms of Fourier series,
\begin{equation}\eta(y)=A+ B(1-y) y+ \sum\limits_{n=1}^{N_f}\left[a_n\cos(k_ny)+b_n\sin(k_ny)\right],
\label{series}
\end{equation}
where $k_n=2\pi n$ and $A,B, a_n, b_n$ are a set of $2N_f+2$ unknown coefficients. To obtain the coefficients $a_n, b_n$ we substitute (\ref{series}) into \eqref{meniscus} and solve the resulting system of linear equations using the collocation method. Constants $A$ and $B$ are then  found from the conditions (\ref{pinning}):
$A=-\sum a_n,\; B=-6(A+1).$


\begin{figure}
\centering
\includegraphics[height=4cm]{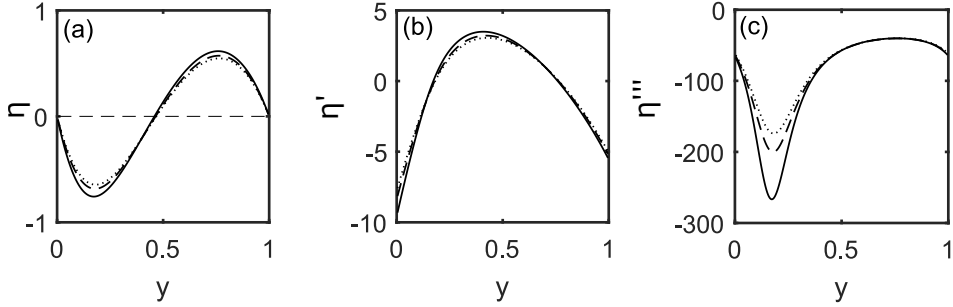}
    \caption{(a) Meniscus shape, (b) its first and (c) third derivatives calculated using $\eps/\mu \ll 1$ and $\Ca/\eps^2 =41.3,\;42.3,\;43.3$  (dotted, dashed and solid lines). }
  \label{critical_profiles}
\end{figure}

If $\eps$ is very small, the solution becomes singular when $\Ca$ approaches some critical value, which is also quite small. Namely, the smallest local thickness of a lubricant film tends to zero, and simultaneously the pressure gradient diverges. This is illustrated in Fig.~\ref{critical_profiles}. We see that the smallest thickness of the lubricant film slightly decreases with the small increase in $\Ca/\eps^2$, but the third derivative of $\eta(y)$, which reflects the growth of pressure gradient in the lubricant neck, changes significantly. Beyond some critical $\Ca$, the solution of positive film thickness does not exist, and
our numerical scheme fails to converge.

\section{Validation of the model based on a lubrication approximation}
\label{valid}

In our asymptotic model we describe the flow inside the groove using the lubrication
theory. It is of considerable interest to determine its accuracy and the range of validity for our configuration.

\begin{figure}
\centering
\includegraphics[height=4cm]{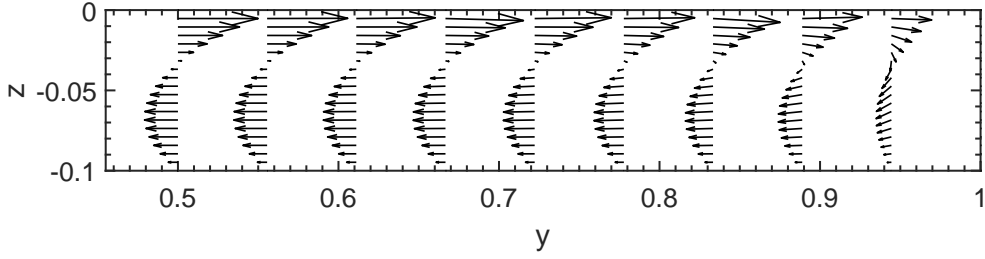}
    \caption{Lubricant velocity field computed using $\varepsilon=0.1$ and $\mu=0.2$. }
  \label{vel_lubr}
\end{figure}

To validate this  approach, here we first present some exact results for a two-phase system with a flat meniscus. Our numerical calculations are based on the method developed by
\citet{ng.co:2010,nizkaya.tv:2014}. Fig. \ref{vel_lubr} shows the lubricant velocity field computed using $\varepsilon =0.1$ and $\mu =0.2$. We see that the velocity field far from
the side texture wall is unidirectional, with a parabolic profile of zero mean flux, confirming all the features described by Eq.\eqref{v_l}. However, in the vicinity of the side wall there is a discernible vertical velocity component, indicating that a simple lubrication model can only be considered as a first approximation.

\begin{figure}
\centering
\includegraphics[height=4cm]{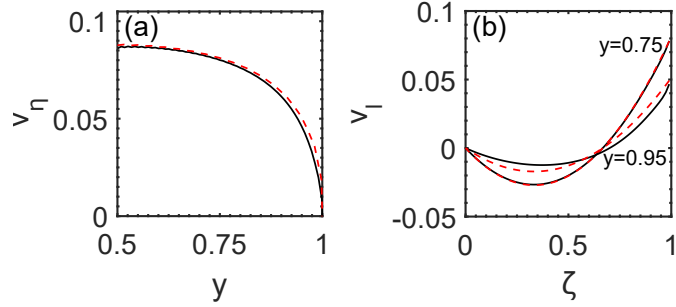}
    \caption{(a) Velocity at the interface of a flat meniscus and (b) inner velocity profiles at cross-sections  $y = 0.75$ and 0.95 calculated for $\eps=0.1$ and $\mu=0.2$. Solid curves show the exact solutions, dashed curves plot results of the lubrication theory. }
  \label{lubr_profiles}
\end{figure}

To examine a significance of these deviations from the lubrication model more
closely, in Fig. \ref{lubr_profiles}(a) we compare the exact numerical results for a velocity at the flat liquid/lubricant interface with predictions of our asymptotic model, i.e. with the solution of \eqref{meniscus} obtained using $\eta=0$. 
The agreement is quite good, but at $y \to 1$ there is some discrepancy, and the lubrication theory slightly overestimates the interface velocity. We have also calculated streamwise velocity profiles in a lubricant. Results for cross-sections $y = 0.75$ and $y = 0.95$ are plotted in Fig. \ref{lubr_profiles}(b). The exact velocity profiles fully coincide with the lubrication model when $y = 0.75$, but close to the side texture wall, $y=0.95$,  we again observe some deviations of the lubrication theory data from the exact results. Nevertheless, the calculations
demonstrate that this discrepancy is small.

Another possible source of inaccuracy of the lubrication model is the use of the first-order approximation for the interface curvature, $\kappa\simeq\eps\eta''$, in Eq.(\ref{norm_stress}), instead of an exact equation for the curvature,
\begin{equation}
\kappa=\dfrac{\eps\eta''}{\left(1+\eps^2\eta'^2\right)^{3/2}}.
\label{ex_cur}
\end{equation}
Eq.\eqref{meniscus} governing the meniscus shape can be rewritten using \eqref{ex_cur} as
\begin{equation}
\left[\frac{\eta ^{\prime \prime }}{\left( 1+\varepsilon ^{2}\eta ^{\prime
}{}^{2}\right) ^{3/2}}\right]^{\prime} =-\frac{6\mu \mathrm{Ca%
}}{\varepsilon ^{3}}\dfrac{v_{\eta }(y)}{(1+\eta )^{2}}.  \label{meniscus1}
\end{equation}
In Section~\ref{res}, we compare the numerical solutions of Eqs.\eqref{meniscus} and \eqref{meniscus1} for several $\mu$ (see Fig.~\ref{diagram}(a) and (b)), and show that they are very close. This implies that $\varepsilon \eta ^{\prime}$ always remains small and can safely be neglected.


\bibliographystyle{jfm}
\bibliography{meniscusbib,viscosity}

\end{document}